# Speculation and Power Law

Sabiou Inoua
inouasabiou@gmail.com
Dec. 2016

(Preliminary version)

**Abstract**

It is now well established empirically that financial price changes are distributed according to a power law, with cubic exponent. This is a fascinating regularity, as it holds for various classes of securities, on various markets, and on various time scales. The universality of this law suggests that there must be some basic, general and stable mechanism behind it. The standard (neoclassical) paradigm implies no such mechanism. Agent-based models of financial markets, on the other hand, exhibit realistic price changes, but they involve relatively complicated, and often mathematically intractable, mechanisms. This paper identifies a simple principle behind the power law: the feedback intrinsic to the very idea of speculation, namely buying when one expects a price rise (and selling when one expects a price fall). By this feedback, price changes follow a random coefficient autoregressive process, and therefore they have a power law by Kesten theorem.

## 1. Introduction

It is well established empirically that financial price changes are distributed according to a power law, with an exponent close to 3. This is a fascinating regularity as it holds for different classes of securities, on various markets, and on various time scales [1, 2]. The dominant neoclassical paradigm doesn't imply anything of this sort, and some researchers think this limitation is due to the failure of this theory to treat the complexity of financial markets, notably the heterogeneity and interactions of financial agents; hence their recourse to another framework, agent-based modeling, in which this complexity is handled computationally (so this is actually a computer-based modeling). Their models exhibit realistic price changes; but they are often complicated and involve so many mechanisms (and parameters) that the key ingredients behind their emergent patterns is difficult to single out [3]. Yet the universality of the empirical power law suggests that there must be some basic mechanism behind it. This paper identifies this mechanism to be speculation itself, in that it induces a feedback in price change. By this feedback, the price change follows a random coefficient autoregressive process, also known as a Kesten process, and so they have a power law by Kesten theorem.

This is a fundamental theorem in the context of financial fluctuations, more generally. For instance, GARCH processes are Kesten processes. But despite their popularity, these models, when fitted to real data, imply an infinite-variance price fluctuation: a power law with exponent two, instead of three [4, 5]. Besides, as is often pointed out, these models are not grounded in a theory of financial markets, but are merely statistical models of the data. Kesten's theorem has also been applied to the study of 'rational bubbles', which are Kesten processes when the discount factor is random [6]. Likewise, however, 'rational bubbles' have a power law with an exponent smaller than unity, which is even more extreme. In certain agent-based models, also, a Kesten process appears as an approximation to a complicated mechanism [7]. But in this paper, price changes follow a Kesten process almost by definition of speculation, making the theory presented here one of the simplest explanation of the financial power law.

The main result of this paper is that the return has a power law with exponent 1 when the speculators' expectations are overall accurate, below 1 when they tend to underestimate the actual price changes, and above 1 when they tend to overestimate it. So only the last case corresponds to the data.

## 2. The empirical law

Let $P_t$ be the price of a financial asset at the closing of period $t$, and let its return (or relative price change) during this period be $r_t \equiv (P_t - P_{t-1})/P_{t-1}$. It is known that for many assets, the return has a power law [1, 2]. That is, as $x$ gets bigger,

$$P(|r|>x) \sim Cx^{-\mu}, \tag{1}$$

where $\mu \approx 3$, and $C > 0$ (the notation means that $P(|r|>x)/Cx^{-\mu} \to 1$ as $x \to \infty$). A second universal property is that, while the return itself is serially uncorrelated, its amplitude (or absolute return) is long-range correlated, a phenomena known as 'volatility clustering'. Fig. 1 shows these regularities for the daily NYSE composite index.

FIG. 1. NYSE composite daily index: (a) Price; (b) Return (in percentage); (c) Tail distribution of absolute return in log-log scale, showing a clear linear decay, and the slope of the least-square fit for absolute returns larger than 2% is -3; (d) The autocorrelation function of return is nearly zero at all nonzero lags, while that of absolute return is nonzero over a long range of lags (volatility clustering).

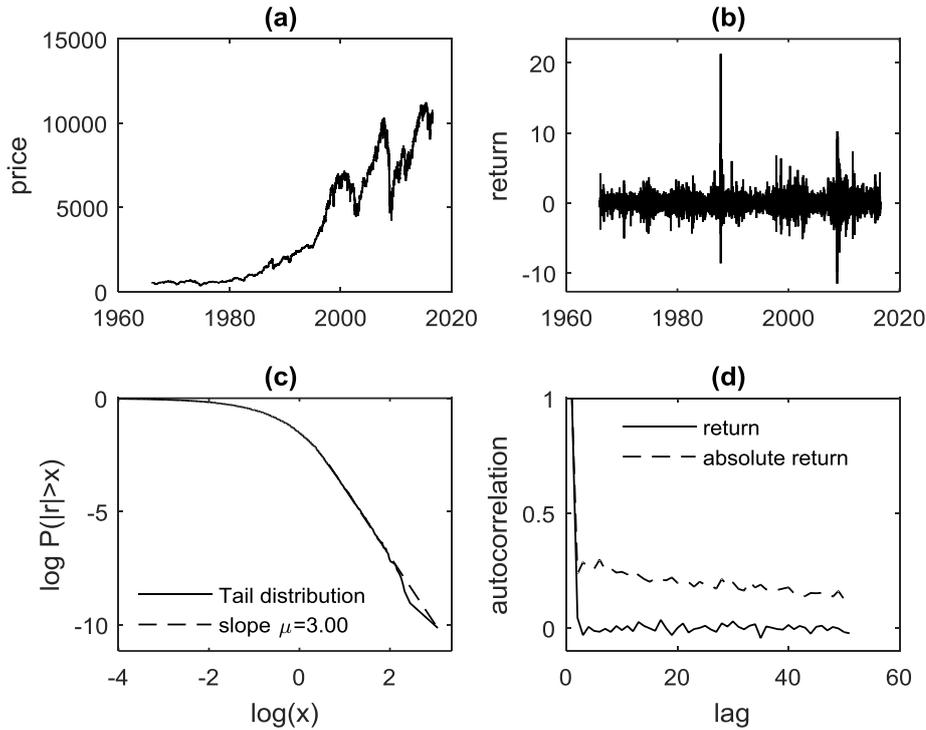

We develop a simple model of speculation that explains the power law and discuss the second property later on.

## 3. A simple theory

Let $x_{it}$ be the signed demand of trader $i$ for a security in period $t$ (that is, $x_{it} > 0$ and $x_{it} < 0$ correspond resp. to a demand and a supply of $|x_{it}|$ units of the asset). In financial markets, supply and demand are expressed by order flows, which are requests to buy or sell a security immediately at the best available price (market orders) or at a given price or better (limit orders). We assume throughout a purely speculative financial market. A *speculator*, by definition, is someone who buys when he expects a price rise, for a future capital gain, and sells when he expects a price fall, to avoid a capital loss. Thus assume demand is proportional to the expected price change, namely

$$x_{it} = \alpha r_{it}, \tag{2}$$

where $r_{it}$ denotes the return that speculator $i$ expects to occur in period $t$, and $\alpha > 0$. Let $N_t$ be the number of speculators who submit orders for the security during $t$ and assume $E(N_t)<\infty$. Excess demand is simply $x_t \equiv \sum_{i=1}^{N_t} x_{it}$, or, given (2), $x_t = \alpha N_t \bar{r}_t$, where $\bar{r}_t \equiv N_t^{-1}\sum_i r_{it}$. Assume the price adjusts proportionally to excess demand, that is,

$$r_t = \beta x_t, \qquad (3)$$

where $\beta > 0$. The two assumptions (2) and (3) combined imply that $r_t = (\alpha\beta N_t)\bar{r}_t$. Finally, assume the speculators are chartists (or trend followers), namely that they use past returns to predict the future return.

All in all, the return is governed by a basic feedback:

$$r_t = a_t \bar{r}_t, \qquad (4)$$

$$\bar{r}_t = F(r_{t-1},...,r_{t-K}), \qquad (5)$$

where

$$a_t = \alpha\beta N_t, \qquad (6)$$

$F$ is a function summarizing the techniques speculators' use to predict the future return from past return data, and $K$ corresponds to the furthest past that speculators overall consider. The key variable behind price changes will turn out to be $a_t$, and hence the number of speculators $N_t$. Theoretically, it simply represents, as (6) implies, the ratio between the actual return and its overall expectation by speculators (a ratio between reality and its perception, so to speak):

$$a_t = r_t / \bar{r}_t. \qquad (7)$$

We distinguish three cases:

(A) Expectations are overall accurate, in the sense that $E(a_t)=1$, or in the sense that $E(\bar{r}_t) = E(r_t)$ and $a_t$ assumes values near 1.

(B) Expectations tend to underrepresent reality on average, in the sense that $E(a_t)>1$.

(C) Expectations tend to overvalue reality, in the sense that $E(a_t)<1$ (which is the case, e.g., when speculators are overconfident by euphoria and over-pessimistic by panic).

The next two sections establish the general result of this paper, namely that the return has a power law with exponent $\mu=1$ in case (A), $\mu<1$ in (B), and $\mu>1$ in (C).

The first case covers 'rational expectation', in the weak sense that the predicted return is an unbiased image of the actual return; we start with it.

### 4. Unbiased expectations

The speculators' return predictions are overall accurate when for instance the individual prediction errors are independent and offset on average. Formally, let these errors be

$$\varepsilon_{it} = r_{it} - r_t, \qquad (8)$$

and assume (i) $E(\varepsilon_{it})=0$, (ii) $\varepsilon_{it}$ and $\varepsilon_{jt+h}$ are independent if $h\neq 0$ or $i\neq j$, and (iii) $E(\varepsilon_{it}^2)<\infty$. The mean expected return is then $\bar{r}_t = r_t + \bar{\varepsilon}_t$, where $\bar{\varepsilon}_t = N_t^{-1}\sum_i \varepsilon_{it}$. When the market is always in equilibrium, namely $x_t = 0$, and the number of speculators $N_t$ is always very large, then it is easy to show that the return is Gaussian by the central limit theorem, which is at odds with the empirical law. But assuming (3) instead of market clearing, that is, assuming the feedback (4)-(5), we get

$$r_t = (1 - a_t)^{-1} e_t, \qquad (9)$$

where $e_t = \alpha\beta \sum_i \varepsilon_{it}$. Given that $E(N_t) < \infty$ and (i)-(iii), we have (iv) $E(a_t) < \infty$, (v) $E(e_t) = 0$, and (vi) $E(e_t^2) < \infty$, which we will use later on.

**Proposition 1**: *The return process* (9) *has a power law with exponent* $\mu = 1$, *as long as* $f_a(1) > 0$, $f_a$ *being the density function of* $a_t$.

Proof: This power law emerges due to the multiplier $(1 - a_t)^{-1}$, which is an inverse: when a random variable can assume values near zero, its inverse is known to have a power law with unit exponent [8, 9]. Indeed for any random variable $X$ with density function $f$ and for any $x > 0$, $P(1/X > x) = P(0 < X < 1/x) = \int_0^{1/x} f(z)dz \sim f(0)/x$ as $x \to \infty$, namely a unit-exponent power law, provided $f(0) > 0$. Similarly, $P(|1/X| > x) \sim 2f(0)/x$. This is one of the simplest ways a power law emerges: amplification by division by near-zeros. Finally, the multiplication by the lighter tailed $e_t$ preserves the power law [10].∎

The process (9) is simulated in Fig. 2.

FIG. 2. Price changes in a speculative market with unbiased expectations: the return process is $r_t = (1 - a_t)^{-1} e_t$. In this simulation, $a_t$ is standard uniform and $e_t$ is standard normal. (a) Dynamics. (b) Tail distribution in log-log scale and least-square fit for values larger than 2%.

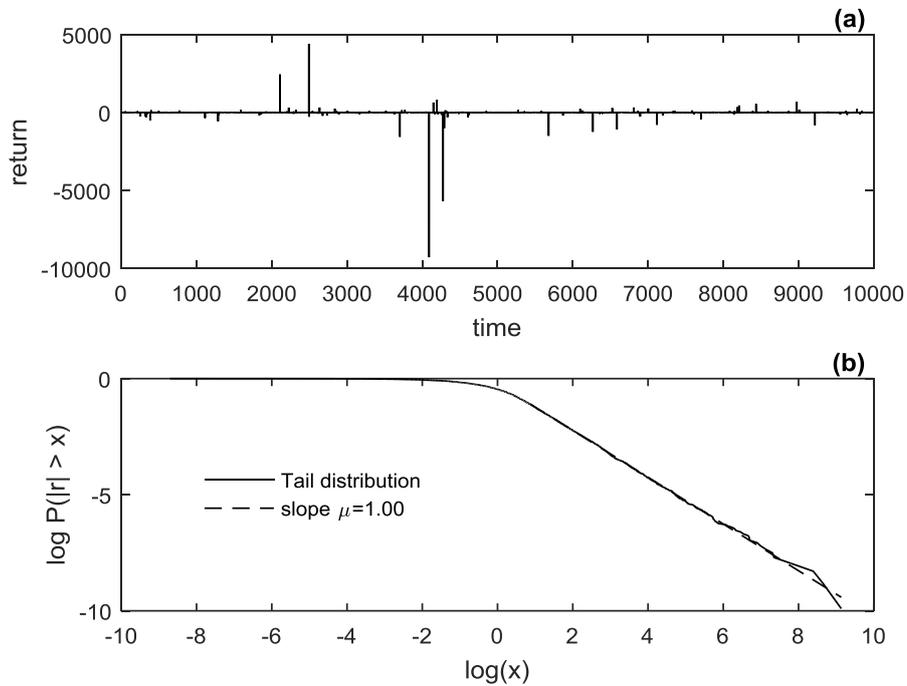

This is an extreme fluctuation: $E(r) = \infty$ for $\mu = 1$ (as is well known and easy to check). So we need a more realistic model of expectations.

## 5. Passive expectations

Assume now the speculators form their opinion of the future return taking the present return as a reference: given $r_{t-1}$, some may expect a higher return in period $t$ while others may expect a lower one; assume these two tendencies offset overall (i.e. speculators overall expect the previous price change to repeat). We refer to this expectation model as 'passive expectations'; it is in a sense the simplest form of trend following; a more general and realistic version is treated later (section 5). Formally, in place of (8), let

$$\varepsilon_{it} = r_{it} - r_{t-1}, \qquad (10)$$

and assume as previously (i)-(iii). The mean expected return is now $\bar{r}_t = r_{t-1} + \bar{\varepsilon}_t$, where $\bar{\varepsilon}_t = N_t^{-1} \sum_i \varepsilon_{it}$. Thus, given (4), the return evolves now according to

$$r_t = a_t r_{t-1} + e_t, \qquad (11)$$

where again $a_t = \alpha \beta N_t$ and $e_t = \alpha \beta \sum_i \varepsilon_{it}$. This is known as a Kesten process. We assume the sequence $\{(a_t, e_t)\}$ consists of independent copies of some random pair $(a, e)$, and that it is exogenous in the sense that $(a_t, e_t)$ is independent of $r_s$ for $s < t$. The following deep theorem by Kesten (and extended by other mathematicians) holds that this process converges in distribution to a power law under mild conditions [11-13]. But the first issue is that of the existence of a stationary solution to this stochastic recurrence equation, which is treated in part 1 of the theorem (see [14] for the general theory of this process).

**Theorem 1 (Kesten)**

*1. (Unique stationary solution) Assume (a) $E[\log(a)] < 0$ and (b) $E[\max(\log|e|, 0)] < \infty$. Then there is a unique strictly stationary solution to* (11)*; moreover, as $t \to \infty$, $r_t$ converges in distribution to this stationary solution for any arbitrary initial value $r_0$.*

*2. (Convergence to power law) Assume: (c) the realizations of $\log(a)$, when $a \neq 0$, are not all integer multiples of some real number; (d) $(1-a)^{-1} e$ doesn't reduce to a constant; there are $\lambda_0, \lambda_1 > 0$ such that (e) $E(a^{\lambda_0}) < 1$, (f) $E(a^{\lambda_1}) \geq 1$ and (g) $E[a^{\lambda_1} \max\{\log(a), 0\}] < \infty$. Then there is a unique positive $\mu \leq \lambda_1$ such that*

$$E(a^\mu) = 1. \qquad (12)$$

*If in addition we have (h) $E(|e|^\mu) < \infty$, then $r_t$ converges in law to a random variable $r$ that has a power law with exponent $\mu$, namely $P(|r| > x) \sim C x^{-\mu}$, where $C > 0$.*

Equation (12), which is sometimes referred as a *Cramer condition*, is a fundamental equation: it implies that the tail of $r$ is a property of $a$ alone (and not $e$); it requires that $P(a_t > 1) > 0$, which is key to the emergence of the power law, for otherwise the tail may decay exponentially [15]. So it is amplifications once again, namely here multiplications by $a_t > 1$, that generate the power law. Though they may look technical, the conditions of this theorem are in fact quite general. Condition (b) holds given (vi), namely, $E(e^2) < \infty$; more generally, (b), (g) and (h) merely requires that $a$ and $e$ be sufficiently light-tailed; (c) and (d) are trivially met here (there is no reason to assume otherwise); by Jensen's inequality, (e) guarantees (a), namely the key condition for stationarity: if $E(a)<1$, $E[\log(a)] \leq \log[E(a)] < 0,$ and in fact because the logarithmic function being strictly convex and $a$ being non-constant, Jensen's inequality is strict, so even when $E(a)=1$, $E[\log(a)] < \log[E(a)] = 0$; (f) is necessary for (12).

In the following proposition, we assume to hold (c), (d), (f), (g), and (h).

**Proposition 2:** *The return process (11) converges to a power law with exponent (A) $\mu=1$ if $E(a)=1$, (B) $\mu<1$ if $E(a)>1$ but $E[\log(a)]<1,$ and (C) $\mu>1$ if $E(a)<1$.*

Proof: This is a direct consequence of Kesten's theorem: (A) $E(a)=1$ means that (12), the Cramer condition, is solved by $\mu=1$; (B) by Jensen's inequality, $E(a^\mu) \geq [E(a)]^\mu > 1$ for $\mu>1$, so (12) is solved only by some $\mu<1$; (C) likewise, $E(a^\mu) \leq [E(a)]^\mu < 1$ for $0<\mu<1$, so that (12) is solved only by some $\mu>1$. ∎

Case (A) confirms the link between accurate expectations and $\mu=1$. Both (A) and (B), are excluded given that $\mu \approx 3$ empirically. Fig. 3 shows a simulation of the return process (11) using an exponential distribution for $a$, in which case it can be shown that $E(a^\mu) = \Gamma(\mu+1)[E(a)]^\mu$ so that $E(a^3)=1$ for $E(a) \approx 0.55$ ($\Gamma$ denotes the gamma function); $e$ is throughout chosen normal with zero mean (its standard deviation is from now on chosen so that the standard deviation of return is 1% as in the NYSE data).

FIG. 3. Price changes in a speculative market with passive expectations: the return process is $r_t = a_t r_{t-1} + e_t$. In this simulation, $a_t$ is exponential with 0.55 mean and $e_t$ is normal with zero mean and 0.65% standard deviation. (a) Price; (b) Return (in percentage); (c) Tail distribution of absolute return in log-log plot and a least-square fit for its values larger than 2%; (d) Autocorrelation function of return and absolute return.

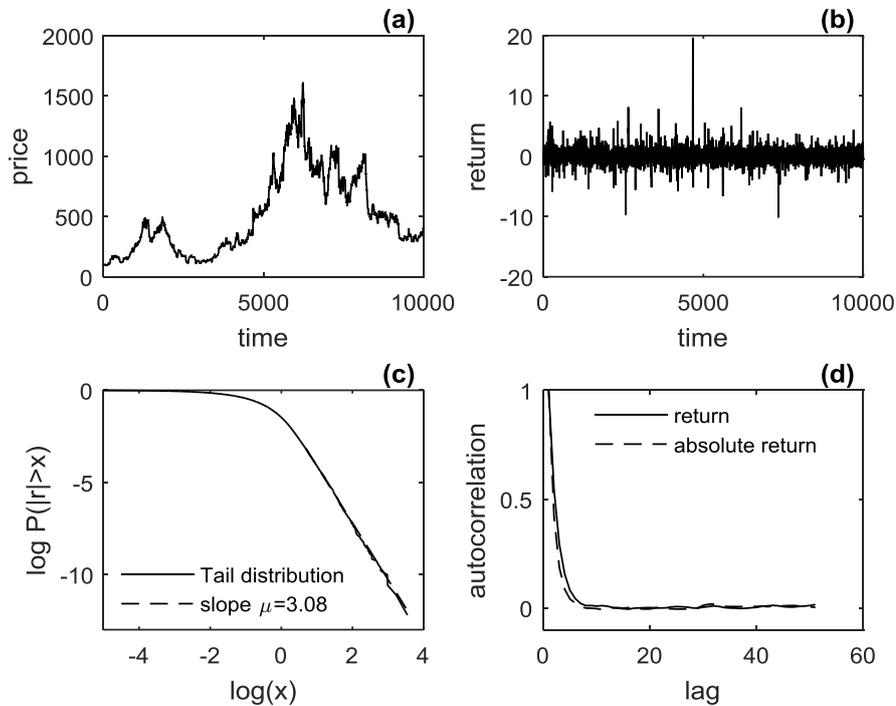

The process (11) thus explains in a natural way the power law of return; it also explains, by the following result, the near absence of serial correlation in the return process.

**Proposition 3:** *The stationary solution $\{r_t\}$ to (11) has an autocorrelation function $\rho(h) \equiv \mathrm{cor}(r_t, r_{t+h}) = [E(a)]^h$, for all $h > 0$, provided that $E(r_t^2) < \infty$.*

Proof: By definition of $\rho(h)$, $E(r_t^2) < \infty$; as $E(e_t) = 0$, $E(r_t) = E(a_t)E(r_{t-1}) = E(a_t)E(r_t)$, i.e. $E(r_t) = 0$. Let $\gamma(h) \equiv E(r_{t+h} r_t)$; $r_{t+h} r_t = a_{t+h} r_{t+h-1} r_t + e_{t+h} r_t$, so $E(r_{t+h} r_t) = E(a_{t+h}) E(r_{t+h-1} r_t)$, i.e. $\gamma(h) = E(a)\gamma(h-1) = [E(a)]^h \gamma(0)$, and $\rho(h) \equiv \gamma(h)/\gamma(0) = [E(a)]^h$. ∎

**Remark 2**: In general, $E(r_t^2) < \infty$ if $E(a^2) < 1$ and $E(e^2) < \infty$ [12]. Also, $E(a^2) < 1$ implies that $E(a) < 1$. Under the conditions of Kesten's theorem, $E(r_t^2) < \infty$ implies that $\mu > 2$.

But the following theorem by Basrak et al. [16] highlights the limitation of this model: it cannot explain volatility clustering!

**Theorem 2**: *Assume $E(a^\varepsilon) < 1$ and $E(|e^\varepsilon|) < \infty$ for some $\varepsilon > 0$. Then the return process (11) converges to its strictly stationary solution $\{r_t\}$, which constitutes a Markov chain. If this chain is irreducible then $\{r_t\}$ is strongly mixing with geometric rate.*

Proof: see Basrak et al. [16, 14]. ∎

The conditions of this theorem are met here given (vi), namely $E(e^2) < \infty$, and the fact that $E(a) < 1$, in the realistic case. This technical notion of 'strong mixing with geometric rate' simply means that any dependence between the present and the future of the process vanishes exponentially. In particular, for any measurable function $f$, $\mathrm{cov}[f(r_{t+h}), f(r_t)]$, if well-defined, decays exponentially with $h$. For instance, $\mathrm{cov}(|r_t|, |r_{t+h}|)$ decays exponentially, as can be seen in Fig. 3 (d); this is at odds with the empirical data (Fig. 1 (d)). So this model cannot account for volatility clustering, namely the high persistence in the amplitude of return, whether measured by absolute or square return.

GARCH processes suffer the same limitation, contrary to a widely held view, as Mikosch and Starica show [4, 5]. Indeed no GARCH process that is realistic in terms of $\mu$ can explain volatility clustering by Theorem 2 [16]. But, again, GARCH processes, when calibrated to fit real data, are not realistic, as they imply $\mu = 2$ and thus $E(r_t^2) = \infty$, so that the autocorrelation is not even defined here. For instance, when the GARCH (1,1) with iid standard normal noise $\{z_t\}$ is fitted to the above-pictured NYSE data, it corresponds to the process $\{r_t\} = \{\sigma_t z_t\}$ where $\{\sigma_t^2\}$ is a Kesten process with $a_t \approx 0.9 + 0.09 z_t^2$ and $e_t \approx 0.01$. As $E(a_t) = 0.99 \approx 1$, $\{\sigma_t^2\}$ is close to a process with $\mu = 1$ and $\{\sigma_t\}$, and hence $\{r_t\}$, to a process with $\mu = 2$, at odds with the empirical $\mu \approx 3$. Only, if we consider the inequality $0.99 < 1$ to be statistically significant enough that $E(r_t^2) < \infty$, then we should consider the autocorrelation of $\{r_t^2\}$, which is the same as that of $\{\sigma_t^2\}$, to decay as $(0.99)^h$, by Proposition 3, which is a relatively slow decay. In reality, it seems that volatility clustering has to do with some non-stationary component in price dynamics [4, 5]. And it is this non-stationary component that fitted GARCH processes come close to mimicking: for $a_t = 0.9 + 0.1 z_t^2$, $E[\log(a_t)] \approx -0.008$, which is negative and thus implies stationarity, but close to zero, and thus close to non-stationarity (see Theorem 1).

The following is a more general and more realistic model in terms of expectations, but shares the same limitation concerning volatility clustering.

## 6. Trend following

We assume more generally that the speculators are chartist in that they use some averages of past returns to predict the future return, that is,

$$r_{it} = \sum_{k=1}^{K} \omega_{ikt} r_{t-k} + \varepsilon_{it}, \qquad (13)$$

where again $K$ corresponds to the furthest past considered relevant by speculators overall, $\varepsilon_{it}$ are exogenous influences on expectations, about which we assume (i)-(iii), and $\sum_{k=1}^{K} \omega_{ikt} = 1$ (speculators may use simple, moving, exponential, or whatever kind of averages). The mean predicted return is $\bar{r}_t = \sum_{k=1}^{K} \bar{\omega}_{kt} r_{t-k} + \bar{\varepsilon}_t$, where $\bar{\omega}_{kt} \equiv N_t^{-1} \sum_{i=1}^{N_t} \omega_{ikt}$ and $\bar{\varepsilon}_t = N_t^{-1} \sum_i \varepsilon_{it}$. So, given (4), the return process is now

$$r_t = a_t \sum_{k=1}^{K} \bar{\omega}_{kt} r_{t-k} + e_t, \qquad (14)$$

where again $e_t = \alpha\beta \sum_i \varepsilon_{it}$. This is a random-coefficient autoregressive process of order $K$, and is also a Kesten process [17]. Indeed, if we let $R_t = [r_t, ..., r_{t-K+1}]^T$, then (14) can be written in matrix form as $R_t = A_t R_{t-1} + E_t$, where $E_t = [e_t, 0, ..., 0]^T$, and

$$A_t = \begin{pmatrix} a_t \bar{\omega}_{1t} & \cdots & a_t \bar{\omega}_{Kt} \\ I_{K-1} & \mathbf{0} \end{pmatrix}, \qquad (15)$$

where $I_{K-1}$ is the identity matrix of order $K-1$ and the bold zero stands for a column of zeros. Kesten's theorem holds for this multidimensional case (in fact it was originally formulated for the general multidimensional case). Given (iv) and (vi), namely $E(a_t) < \infty$ and $E(e_t^2) < \infty$, a sufficient condition for the existence of a unique strictly stationary solution to (14) is that $E(\log\|A_t\|) < 0$, for some arbitrary norm [18, 17, 14].

**Proposition 5**: *When $E[\log(a_t)] < 0$, the return process (14) converges in distribution to a strictly stationary process.*

Proof: $\|A_t\|_\infty = \max\{1, a_t\}$, given that $\sum_{k=1}^{K} \bar{\omega}_{kt} = 1$; so $E(\log\|A_t\|_\infty) = P(a_t > 1)E[\log(a_t)]$. ∎

It is a deep property that stationarity depends essentially on $a_t$ alone, and not on the details of how expectations are formed, namely on the individual weights. More generally, all that is said about the previous simplified model holds for this generalized model as well. Notably, the limiting distribution of return is a power law with an exponent given now by

$$\lim_{t \to \infty} \frac{1}{t} \log(E\|A_1 \cdots A_t\|^\mu) = 0, \qquad (16)$$

which is a generalization of (12), but which is much more difficult to solve, except via computer simulations. Qualitatively, as we said, the process (13) is equivalent to its simplified version (11): simulations suggest that Proposition 2 holds for this general formulation (when $E(a) = 1$, e.g., $\mu = 1$). It is a good exercise to try to prove this result.

Fig. 4 shows a realistic simulation of the process (14).

FIG. 4. Price changes in a market of trend-following speculators: the return process is $r_t = a_t \sum_{k=1}^{K} \bar{\omega}_{kt} r_{t-k} + e_t$. Here $a_t$ is exponential with mean 0.6; $K = 3$, $\bar{\omega}_{1t}, \bar{\omega}_{2t}, \bar{\omega}_{3t}$ are uniformly distributed between 0.7 and 0.8, 0.1 and 0.2, and 0 and 0.2, resp.; and $e_t$ is normal with zero mean and 0.7% standard deviation. (a) Price; (b) Return; (c) Tail distribution of absolute return and least-square fit for values larger than 2%; (d) Autocorrelation function of

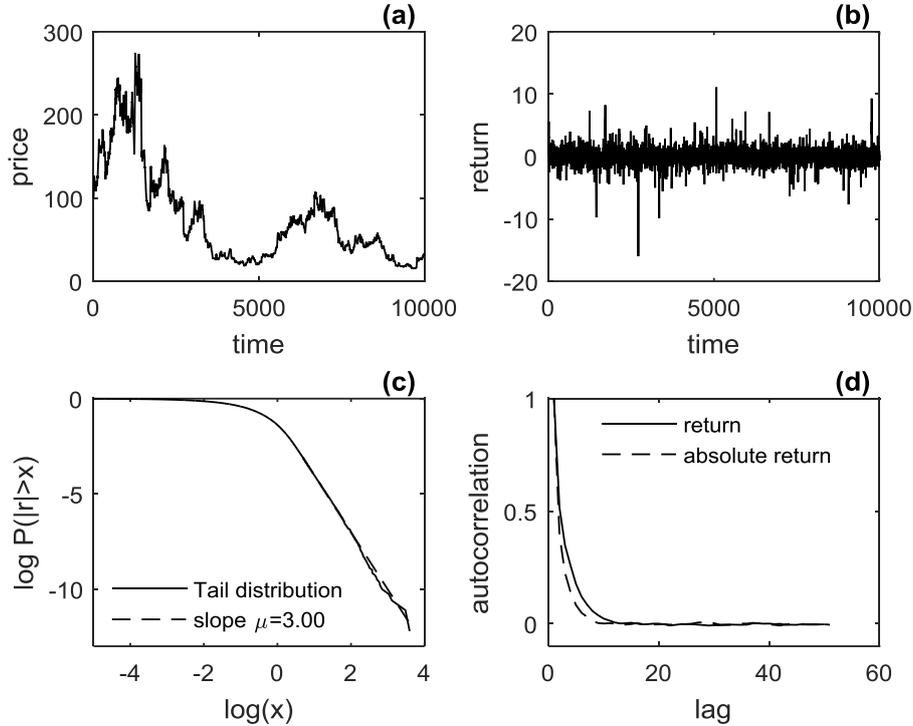

## 7. Conclusion

In sum, there is an intimate link between speculation and the power law of financial price changes. The fundamental variable behind this fluctuation is the ratio between the actual and the overall expected returns, which turns out to be proportional to the number of speculators transacting on the security. On this point also, this theory departs from most agent-based models, in which the number of agents is fixed. Intuitively, the distribution of this variable is most likely similar for most securities: fundamentally, how expectations are formed is a property of human mind and is not likely to change dramatically in time and with the object of the speculation, namely the security. This might explain the universality of the power law. An essential limitation of this theory, however, is that it cannot explain volatility clustering, a phenomenon for which we should also seek a basic underlying mechanism. This is the subject of a forthcoming paper.